\documentclass[prd,amsmath,amssymb,showpacs,superscriptaddress,nofootinbib,twocolumn]{revtex4-1}
\usepackage{graphicx}
\usepackage{epsfig}
\usepackage{psfrag,amssymb}
\usepackage{lineno}
\usepackage{dcolumn}
\usepackage{bm}
\usepackage{xspace}
\usepackage{xcolor}
\usepackage{rotating}
\usepackage{multirow}
\usepackage{colortbl}
\usepackage{epstopdf}
\usepackage{makecell}

\usepackage[colorlinks=true,linkcolor=blue,anchorcolor=blue,citecolor=blue]{hyperref}

\newcommand{\etap}{\eta^{\prime}}

\newcommand{\psipipieta}{\psi(3686)\to \pi^+\pi^-\eta}

\newcommand{\BESIIIorcid}[1]{\href{https://orcid.org/#1}{\hspace*{0.1em}\raisebox{-0.45ex}{\includegraphics[width=1em]{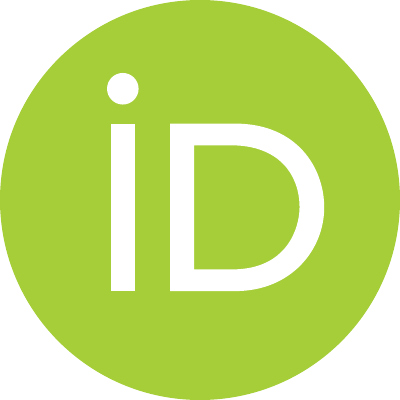}}}}

\newcommand{\BESIII}{BES\uppercase\expandafter{\romannumeral3}\xspace}

\begin{document}
\makeatletter
\def\@citecolor{blue}
\def\@linkcolor{blue}
\def\@anchorcolor{blue}
\def\@urlcolor{blue}
\makeatother

\title{\boldmath Improved measurement of $\psi(3686)\to \pi^+\pi^- \eta$ }
\author{
\begin{small}
  \begin{center}
M.~Ablikim$^{1}$\BESIIIorcid{0000-0002-3935-619X},
M.~N.~Achasov$^{4,c}$\BESIIIorcid{0000-0002-9400-8622},
P.~Adlarson$^{84}$\BESIIIorcid{0000-0001-6280-3851},
X.~C.~Ai$^{90}$\BESIIIorcid{0000-0003-3856-2415},
C.~S.~Akondi$^{32A,32B}$\BESIIIorcid{0000-0001-6303-5217},
R.~Aliberti$^{40}$\BESIIIorcid{0000-0003-3500-4012},
A.~Amoroso$^{83A,83C}$\BESIIIorcid{0000-0002-3095-8610},
Q.~An$^{79,66,\dagger}$,
Y.~H.~An$^{90}$\BESIIIorcid{0009-0008-3419-0849},
M.~S.~Anderson$^{40}$\BESIIIorcid{0009-0008-1550-2632}
% ...
\\
\vspace{0.2cm}
(BESIII Collaboration)\\
\vspace{0.2cm} {\it
$^{1}$ Institute of High Energy Physics, Beijing 100049, People's Republic of China\\
% ...
$^{4}$ Budker Institute of Nuclear Physics SB RAS (BINP), Novosibirsk 630090, Russia\\
% ...
$^{32}$ INFN Sezione di Ferrara, (A)INFN Sezione di Ferrara, I-44122, Ferrara, Italy; (B)University of Ferrara, I-44122, Ferrara, Italy\\
% ...
$^{40}$ Johannes Gutenberg University of Mainz, Johann-Joachim-Becher-Weg 45, D-55099 Mainz, Germany\\
% ...
$^{66}$ State Key Laboratory of Particle Detection and Electronics, Beijing 100049, Hefei 230026, People's Republic of China\\
% ...
$^{79}$ University of Science and Technology of China, Hefei 230026, People's Republic of China\\
% ...
$^{83}$ University of Turin and INFN, (A)University of Turin, I-10125, Turin, Italy; (B)University of Eastern Piedmont, I-15121, Alessandria, Italy; (C)INFN, I-10125, Turin, Italy\\
$^{84}$ Uppsala University, Box 516, SE-75120 Uppsala, Sweden\\
% ...
$^{90}$ Zhengzhou University, Zhengzhou 450001, People's Republic of China\\

\vspace{0.2cm}
$^{\dagger}$ Deceased\\
% ...
$^{c}$ Also at the Novosibirsk State University, Novosibirsk, 630090, Russia\\
% ...

}

\end{center}
\end{small}
}

\date{\today}

\begin{abstract}
Using a sample of $(2712.4 \pm 14.3) \times 10^6$ $\psi(3686)$ events collected with the BESIII detector, we perform a partial wave analysis of the  decay $\psi(3686)\to\pi^{+}\pi^{-}\eta$  with the $\pi^{+}\pi^{-}$ invariant mass in the $\rho(770)^0$ mass region, [0.3, 2.2]~$\mathrm{GeV}/c^2$. The cross section line-shape around the $\psi(3686)$ resonance is measured using energy-scan data sets, enabling the estimation of continuum contribution and its interference with the $\psi(3686)$ resonance. The branching fraction of $\psi(3686)\to\rho(770)^0\eta$ is determined to be $(1.34 \pm 0.03\pm0.13) \times 10^{-5}$. The  $\psi(3686)\to a_{2}(1320)^{\pm}\pi^{\mp}$ decay is observed for the first time with a significance of $35.7\sigma$, and its 
partial branching fraction is measured to be $(2.13 \pm 0.05\pm0.32) \times 10^{-5}$.  The $\eta$-$\eta^{\prime}$ pseudoscalar mixing angle $\theta_P$ is then found to be $(-11.2 \pm 1.6\pm3.6)^\circ$ or $(-19.4\pm 3.1\pm5.7)^\circ$, where the first uncertainty is statistical and the second systematic.
\end{abstract}

\maketitle

\section{INTRODUCTION}

The charmonium state $\psi(3686)$ provides a unique laboratory for studying nonperturbative quantum chromodynamics (QCD). Its decays into vector-pseudoscalar (VP) meson pairs serve as important probes of strong interactions~\cite{Escribano:2009jti,Bramon:1997mf,Thomas:2007uy}. In the $\psi(3686)$ energy region, decay dynamics are influenced by multiple mechanisms including hadronic transitions, potential glueball mixing, and SU(3) flavor symmetry breaking~\cite{Gilman:1987ax,Benayoun:1999fv,Bramon:1997va}. 
Understanding their relative contributions is essential for interpreting the decay mechanisms of $\psi(3686)$.

The isospin-violating decays \(\psi(3686)\to \rho(770)^0\eta\) and \(\psi(3686)\to \rho(770)^0\eta'\) provide sensitive probes of the \(\eta\)-\(\eta'\) mixing angle \(\theta_P\). In the singly OZI-suppressed annihilation framework, \(\eta'\) production requires either substantial gluonic components or pseudoscalar mixing.
Such a constraint offers a unique window to investigate SU(3) flavor symmetry breaking and possible exotic contributions to meson wavefunctions\cite{intr_mix4,intr_scalar}.

 The physical meson states $\eta$ and $\etap$ are related to the flavor basis via the usual mixing foumulae:
\begin{equation}
\begin{pmatrix}
\eta \\ \eta^{\prime}
\end{pmatrix} =
\begin{pmatrix}
\cos\theta_P & -\sin\theta_P \\
\sin\theta_P & \cos\theta_P
\end{pmatrix}
\begin{pmatrix}
\eta_8 \\ \eta_1
\end{pmatrix},
\end{equation}
where  $\theta_P$ is the pseudoscalar mixing angle, $\eta_8 = (u\bar{u} + d\bar{d} - 2 s\bar{s})/\sqrt{6}$ and $\eta_1 = (u\bar{u} + d\bar{d} + s\bar{s})/\sqrt{3}$ are the SU(3) octet and singlet states, respectively.

The $\eta-\eta^{\prime}$ mixing angle $\theta_P$ can also be determined by relating it to the physical meson masses through the diagonalization of the mass matrix. This relationship is given by:
\begin{equation}
\tan^2 \theta_P = \frac{4 m_K - m_\pi - 3 m_{\eta}}{-4 m_K + m_\pi + 3 m_{\eta^{\prime}}},
\label{theta_ge}
\end{equation}
and substituting the pseudoscalar nonet masses~\cite{PDG_2024} yields $\theta_{\text{P,lin}} = -24.5^\circ$, or $\theta_{\text{P,quad}} = -11.3^\circ$ if the massed are instead all replaced by their squares. Lattice QCD calculations give $\theta_P = (-14.1 \pm 2.8)^\circ$~\cite{PDG_2024,Christ:2010dd}.

Based on $J/\psi \to \text{VP}$ decay data, the DM2 Collaboration extracts the  mixing angle $\theta_P = (-19.1\pm 1.4)^\circ$~\cite{DM2:1988bfq}.
The MARKIII Collaboration also reported $\theta_P = (-19.2\pm 1.4)^\circ$ using $J/\psi \to \text{VP}$ decays~\cite{MarkIII:1988pvy}. The KLOE Collaboration measured the ratio of radiative decay widths $\Gamma(\phi \to \gamma\eta^{\prime})/\Gamma(\phi \to \gamma\eta)$ and derived $\theta_P = (-12.9^{+1.9}_{-1.6})^\circ$~\cite{KLOE:2002jed}. 
The BESII  Collaboration measured $J/\psi \to \gamma\pi^0,\gamma\eta,\gamma\eta'$ and obtained $\theta_P = (-22.08 \pm 0.81)^\circ$~\cite{BES:2005otq}. 
The BESIII Collaboration performed improved measurements of the same processes to determine $\theta_P = (-22.11 \pm 0.26)^\circ$ and $(-19.34 \pm 0.34)^\circ$, for two different phenomenological models~\cite{BESIII:2023fai}.

In the absence of gluonium contributions to  $\eta$-$\eta^{\prime}$ mixing, the mixing angle $\theta_P$ can also be determined by comparing the reduced branching fractions of $\psi(3686)\to\rho(770)^0\eta$ and $\psi(3686)\to\rho(770)^0\eta^{\prime}$ decays~\cite{Bramon_1997}:
\begin{equation}
R_P = \frac{\tilde{\mathcal{B}}(\psi(3686)\to\rho(770)^0\eta)}{\tilde{\mathcal{B}}(\psi(3686)\to\rho(770)^0\eta^{\prime})} = \tan^2(\theta_I - \theta_P),
\label{Ratio}
\end{equation}
where $\theta_I = 35.3^\circ$ is the ideal mixing angle, the reduced branching ratio $\tilde{\mathcal{B}}(\psi \to \text{VP}) \equiv \mathcal{B}(\psi \to \text{VP})/p_V^3$, with $p_V$ being the vector meson momentum in the $\psi(3686)$ rest frame.

Experimentally, the decay $\psi(3686) \to \rho(770)^0\eta$ was first measured by the BES collaboration based on $14.0\times 10^6$ $\psi(3686)$ events, yielding $\mathcal{B} = (1.78^{+0.67}_{-0.62}\pm0.17)\times 10^{-5}$~\cite{BES_rhoeta}. Subsequently, the CLEO collaboration performed a measurement based on $3.08\times10^6$ $\psi(3686)$ events, reporting $\mathcal{B} = (3.0^{+1.1}_{-0.9}\pm0.2)\times 10^{-5}$~\cite{CLEO_rhoeta}, which is somewhat higher than the BES result. With the same dataset, CLEO also set an upper limit of $1.6 \times 10^{-4}$ on the branching fraction of $\psi(3686) \to \pi^+\pi^-\eta$~\cite{CLEO_pipieta}. These previous measurements suffer from large uncertainties mainly due to limited statistics.

Using the world's largest $\psi(3686)$ data sample collected with the BESIII detector~\cite{N_all}, we perform a partial wave analysis (PWA) to improve the measurement of $\psi(3686)\to\rho(770)^0\eta$. Interference effects are evaluated by analyzing the line-shape of the cross section for $e^+e^-\to\pi^+\pi^-\eta$ with energy-scan datasets taken near the $\psi(3686)$ resonance. Subsequently, the $\eta$-$\eta^{\prime}$ mixing angle $\theta_P$ is extracted using Eq.~\ref{Ratio} and the BESIII measured value for $\psi(3686)\to\rho(770)^0\eta^{\prime}$~\cite{BESIII_rhoetap}.

\section{BESIII DETECTOR AND DATA SAMPLES}

The BESIII detector~\cite{BESIII_2009_detector} is a magnetic spectrometer operating at the BEPCII $e^+e^-$ collider~\cite{BESIII_2016_detector}. It records symmetric $e^+e^-$ collisions in the center-of-mass energy range from 1.84 to 4.95~GeV, with a peak luminosity of $1.1 \times 10^{33}\;\text{cm}^{-2}\text{s}^{-1}$ achieved at $\sqrt{s} = 3.773$~GeV. The cylindrical core of the BESIII detector covers 93\% of the full solid angle and consists of a helium-based multilayer drift chamber (MDC), a plastic scintillator time-of-flight (TOF) system, and a CsI(Tl) electromagnetic calorimeter (EMC), all enclosed in a superconducting solenoidal magnet providing a 1.0 T magnetic field. The solenoid is supported by an octagonal flux-return yoke with resistive plate counter muon identification modules interleaved with steel.

The charged-particle momentum resolution at 1 GeV/$c$ is 0.5\%, and the resolution of the specific ionization energy loss ($\mathrm{d}E/\mathrm{d}x$) in the MDC  is 6\% for electrons from Bhabha scattering. The EMC measures photon energies with a resolution of 2.5\% (5\%) at 1 GeV in the barrel (end cap) region. The time resolution in the TOF barrel region is 68 ps, while that in the end cap region was 110 ps. The end cap TOF system was upgraded in 2015 using multigap resistive plate chamber technology, providing a time resolution of 60 ps~\cite{BESIII_2019_detector}; about 90\% of the data used in this analysis benefits from this upgrade.

For this analysis, we use datasets collected by the BESIII experiment in 2009, 2012, 2018, and 2021, which include $(2712.4 \pm 14.3) \times 10^6$ $\psi(3686)$ events at the peak energy and data from 11 energy-scan points near the $\psi(3686)$ region~\cite{N_all}, corresponding to a total integrated luminosity of approximately $5.185\ \mathrm{fb}^{-1}$. The center-of-mass energies are measured via a beam energy measurement system using $\text{CO}_2$ laser Compton backscattering~\cite{Abakumova:2011rp}, and the integrated luminosities are determined using $e^+ e^-\to\gamma\gamma$ events~\cite{BESIII:2017lkp}.

Monte Carlo (MC) simulated samples are produced with a GEANT4-based~\cite{GEANT4:2002zbu} software package that includes the geometric description of the BESIII detector and its response. The simulation models the beam energy spread and initial state radiation (ISR) in $e^+e^-$ annihilations using the KKMC generator~\cite{Jadach:2000ir}. Possible backgrounds are studied using an inclusive $\psi(3686)$ MC. The inclusive MC sample includes the production of the $\psi(3686)$ resonance, the ISR production of the $J/\psi$, and the continuum processes incorporated in KKMC. All particle decays are modeled with BESEVTGEN~\cite{Lange:2001uf,Ping:2008zz} using branching fractions from the Particle Data Group (PDG)~\cite{PDG_2024} when available, or otherwise estimated with LUNDCHARM~\cite{Chen:2000tv}. Final state radiation from charged particles is incorporated using the PHOTOS package~\cite{Richter-Was:1992hxq}. The signal process $e^+e^-\to\gamma^* \to \psi(3686)$ is generated with the CONEXC generator~\cite{Ping:2013jka}, and the further decay to $\pi^+\pi^-\eta$ is generated using covariant tensor amplitudes with parameters fixed to the PWA results from this analysis. 

\section{EVENT SELECTION}

Candidates for $e^+e^-\to \pi^+\pi^-\eta$ with $\eta\to\gamma\gamma$ are required to have exactly two charged tracks with opposite charge and at least two photon candidates. Charged tracks are reconstructed from MDC hits and must satisfy $|\cos\theta| < 0.93$, where $\theta$ is the polar angle with respect to the beam direction. The distance of closest approach must be within 10 cm along the beam direction and 1 cm in the transverse plane. 

Particle identification (PID) combines measurements of $\mathrm{d}E/\mathrm{d}x$ in the MDC and time of flight from the TOF. Each track must have a higher probability under the pion hypothesis than under the electron hypothesis, with the pion probability exceeding 0.001.

Photon candidates are identified from isolated electromagnetic showers in the EMC. The energy deposit must be at least 25 MeV in the barrel region ($|\cos\theta| < 0.8$) or 50 MeV in the end cap region ($0.86 < |\cos\theta| < 0.92$). 
To suppress electronic noise and beam-related backgrounds, the timing of the EMC shower must be within [0,700]~ns relative to the event start time.
To exclude showers from charged particles, the opening angle between the photon candidate and any charged track must exceed $10^\circ$.

A four-constraint (4C) kinematic fit is applied under the hypothesis $e^+e^- \to \pi^+\pi^-\gamma\gamma$, constraining the measured four-momenta of all final-state particles to the initial four-momentum of the $e^+e^-$ system. For each event, the $\pi^+\pi^-\gamma\gamma$ combination with the smallest $\chi^2_{\rm 4C}$ is retained. The optimal requirement $\chi^2_{\rm 4C} < 46$ is determined by maximizing the figure of merit $S/\sqrt{S+B}$, where $S$ and $B$ represent the expected signal and background yields, respectively.

Since events of interest are primarily located in the $\rho(770)^0$ mass region, we require that the $\pi^+\pi^-$ invariant mass $M_{\pi^+\pi^-}$ lie in the region [0.3, 2.2]$~\mathrm{GeV}/c^2$.  To suppress backgrounds from $\psi(3686) \to \gamma\chi_{cJ}$, the recoil mass against the lower-energy photon ($\gamma_2$), $M^{\rm rec}_{\gamma_2}$, must be less than 3.47~GeV/$c^2$.

Potential background sources are studied using an inclusive $\psi(3686)$ MC (INC-MC) and exclusive  $\psi(3686) \to  \gamma \chi_{c1} \to \gamma \pi^+ \pi^- \eta$ and $\psi(3686) \to  \gamma \pi^+ \pi^- \eta$ MC samples  after applying the same selection criteria. 
Study of the INC-MC shows that the largest background contribution comes from
 $\psi(3686) \to  \pi^+ \pi^- \eta\gamma, \eta \to \gamma\gamma$ decays. 
Non-$\eta$ backgrounds are estimated using events in the $\eta$ sideband region of the 3.6861 GeV sample, 
defined as $M_{\gamma\gamma} \in [0.475,0.500]~\mathrm{GeV}/c^2$. 

\begin{figure}[htbp]
    \centering
\includegraphics[width=0.42\textwidth]{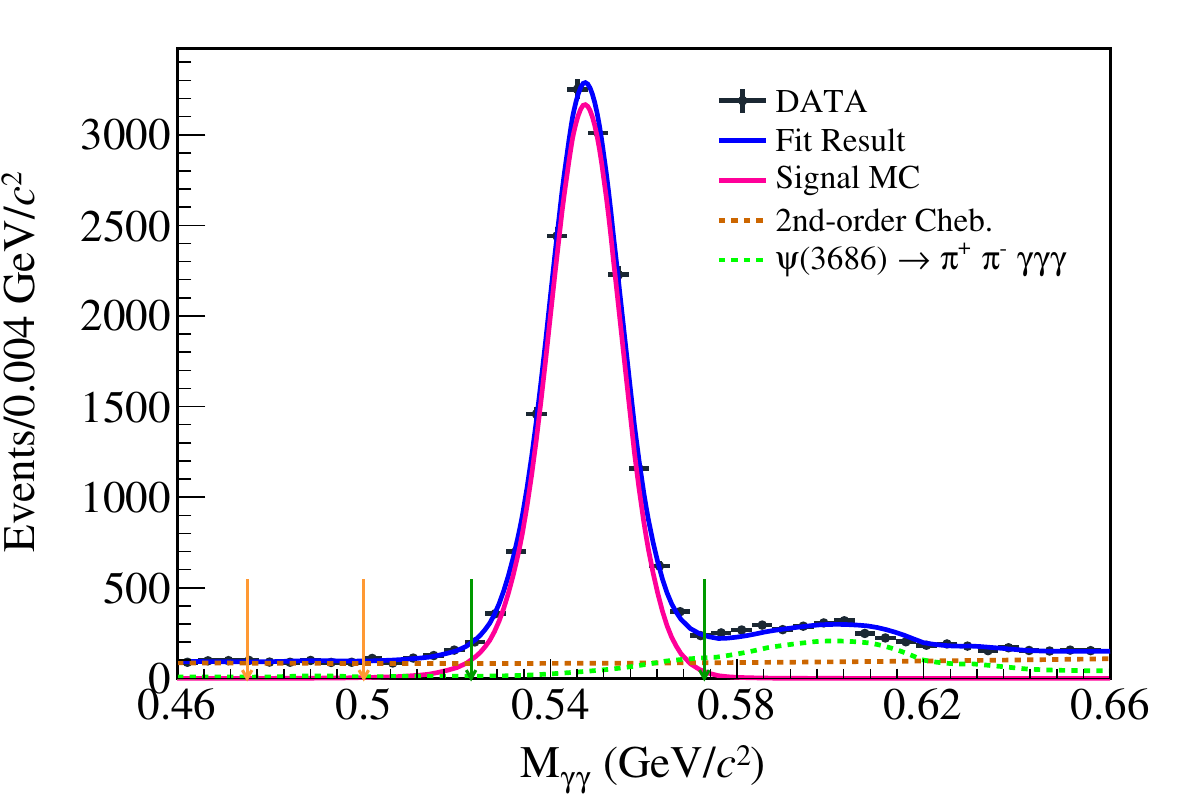}
    \caption{Fit to the $M_{\gamma\gamma}$ distribution at $\sqrt{s} = 3.6861$ GeV. The green arrows indicate the $\eta$ signal region, and the orange arrows indicate the sideband region.}
    \label{fig:etamass}
\end{figure}

The resulting distribution of $\gamma\gamma$ invariant mass, $M_{\gamma\gamma}$, is shown in Fig.~\ref{fig:etamass}. The $\eta$ signal region defined as $M_{\gamma\gamma}\in[0.523, 0.573]\,\mathrm{GeV}/c^{2}$. 
The signal is described
by the MC-simulated shape.
The non-peaking background is described by a second-order Chebyshev polynomial function and the peaking background
described by the INC-MC $\psi(3686)  \to \pi^+ \pi^- \gamma\gamma\gamma$ decays.
The background fraction is 9.5\%. In the PWA, the flat background is subtracted using the sideband and the peaking background by using the INC-MC sample.

\section{PARTIAL WAVE ANALYSIS}

\subsection{Amplitude construction}

A PWA is performed on 15,910 candidate events for $\psi(3686)\to \pi^+\pi^-\eta$ with the $\pi^{+}\pi^{-}$ invariant mass in the $\rho(770)^0$ mass region [0.3, 2.2]~$\mathrm{GeV}/c^2$, with an expected background of 1500 events at $\sqrt{s}=3.6861$~GeV. The decay amplitude is constructed using the covariant tensor amplitude formalism, with the full fit procedure implemented within the FDC-PWA framework~\cite{FDC_2004}. Amplitudes are derived from an effective Lagrangian that respects C- and P-parity conservation and hermiticity.

The $\rho(770)^0$ resonance is described with a Breit-Wigner form incorporating a mass-dependent width~\cite{Ping:2007ia}:
\begin{equation}
    BW_{\rho}(s)=\frac{1}{s-m^{2}_{\rho}+i\sqrt{s} \, \Gamma_{\rho}(s)},
\end{equation}
where
\begin{equation}
    \Gamma_{\rho}(s)=\Gamma_{\rho}\left(m_{\rho}^{2}\right) \, \frac{m_{\rho}^{2}}{s}\left(\frac{q(s)}{q_0}\right)^{3}\frac{B_{1}^{2}(q(s))}{B_{1}^{2}(q_0)}.
  \end{equation}
Here $s$ is the square of the $\pi^+\pi^-$ invariant mass, $q(s)$ is the three-momentum magnitude in the $\pi^+\pi^-$ rest frame, $q_0 = q(m_{\rho}^{2})$, and $B_L$ are Blatt-Weisskopf barrier factors given by:
\begin{align}
B_1(q)/B_1(q_0) &= \frac{\sqrt{1 + (dq_0)^2}}{\sqrt{1 + (dq)^2}}, 
\label{eq:blattweisskopf}
\end{align}
with a fixed $d = 3~\mathrm{GeV}^{-1}$~\cite{LHCb:2015yax}. Other resonances  are described by relativistic Breit-Wigner functions with constant widths:
\begin{equation}
    BW(m)=\frac{1}{m^{2}_{0}-m^{2}-im_{0}\Gamma_0}.
\end{equation}

\subsection{Fit method}

The joint probability density for observing $N$ events is
\begin{equation}
    \mathcal{L} = \prod_{i=1}^{N}P(x_{i}),
\end{equation}
where $P(x_i)$ is the probability to produce event $i$ with four-momentum set $x_i$. The normalized $P(x_i)$ is calculated from the differential cross section:
\begin{equation}
    P(x_{i})=\frac{(d\sigma/d\Phi)_{i}}{\sigma_{\rm MC}},
\end{equation}
where the normalization factor $\sigma_{\rm MC}$ is evaluated using a large MC sample of $N_{\rm MC}$ accepted events generated with a phase-space model and subject to the same selection criteria as data:
\begin{equation}
    \sigma_{\rm MC}=\frac{1}{N_{\rm MC}}\sum_{i=1}^{N_{\rm MC}}\left(\frac{d\sigma}{d\Phi}\right)_{i}.
\end{equation}

We minimize $S=-\ln\mathcal{L}$ using the MINUIT package. Background subtraction is implemented by replacing the likelihood function with:
\begin{equation}
    S = -(\ln\mathcal{L}_{\rm data}-\ln\mathcal{L}_{\rm bkg}),
\end{equation}
where the background is described by normalized sideband events. The signal yield for a given intermediate resonance is estimated by scaling its cross section fraction to the net signal events:
\begin{equation}
    N_{i}=R_{i}(N_{\rm obs}-N_{\rm bkg}),
\end{equation}
where $R_{i}=\sigma_{i}/\sigma_{\rm tot}$ is the ratio of the cross section for resonance $i$ to the total cross section, and $N_{\rm obs}$ and $N_{\rm bkg}$ are the numbers of observed and background events, respectively.

Statistical uncertainties $\delta N_i$ are estimated via the bootstrap method~\cite{Bootstrasp}:
\begin{equation}
  \delta N_i = \sqrt{\frac{1}{n}\sum_{j=1}^{n}
               \left(N_{i,j}-\bar{N}_i\right)^2},
\end{equation}
where $N_{i,j}$ is the yield from the $j^{th}$ bootstrap sample, $\bar{N}_i$ is the average yield over $n=500$ samples.

\subsection{PWA results}

The fit includes all possible intermediate states from the PDG~\cite{PDG_2024} that satisfy quantum number conservation in subsequent two-body decays. 

The PWA fit procedure starts by including the resonances \( \omega \), \( \rho(770)^0 \), \( \rho(1450)^0 \),  \( \rho_3(1690)^0 \), \( \rho(1700)^0\), \( \rho(1900)^0 \) , \( \rho_3(1990)^0 \),  
    \( a_2(1320)^{\pm} \) and \( a_2(1700)^{\pm} \) , as well  
as the nonresonant  \( NR(1^-) \),\( NR(3^-) \) and \( NR(2^+) \) 
components as the initial amplitudes. The masses and widths of resonances are fixed to the PDG values~\cite{PDG_2024}. 
The statistical significance of each component is evaluated from the change in $S$ upon its inclusion or removal, with the change in the number of degrees of freedom taken into account. 

The nominal fit hypothesis is obtained by dropping the least significant component and re-fitting until all components have a statistical significance exceeding $5\sigma$ and yield greater than 1\%.   Those components are the \( \rho(770)^0 \),  \( \rho_3(1690)^0 \), \( a_2(1320)^{\pm} \), \( a_2(1700)^{\pm}\), \( NR(1^-) \). Statistical significances are listed in Table~\ref{tab:significance}. The decay chains are:
\begin{align*}
&\psi(3686)\to\rho(770)^0\eta,\ \rho(770)^0\to\pi^{+}\pi^{-},\\
&\psi(3686)\to\rho_{3}(1690)^0\eta,\ \rho_{3}(1690)^0\to\pi^{+}\pi^{-},\\
&\psi(3686)\to a_{2}(1320)^{\pm}\pi^{\mp},\ a_{2}(1320)^{\pm}\to\pi^{\pm}\eta,\\
&\psi(3686)\to a_{2}(1700)^{\pm}\pi^{\mp},\ a_{2}(1700)^{\pm}\to\pi^{\pm}\eta.
\end{align*}

\begin{table}[htbp]
    \centering
    \caption{Statistical significances of components in the nominal PWA model.}
    \label{tab:significance}
    \begin{tabular}{l c}
    \hline\hline
    Resonance & Significance \\ \hline
    $\rho(770)^0$       & $>36 \sigma$ \\
    $\rho_{3}(1690)^0$  & $22.1\sigma$ \\
    $a_{2}(1320)^{\pm}$ & $35.7\sigma$ \\
    $a_{2}(1700)^{\pm}$ & $23.1\sigma$ \\
    NR($1^{-}$)         & $18.0\sigma$ \\ \hline\hline
    \end{tabular}
\end{table}

\begin{figure*}[htbp]
    \centering    \includegraphics[width=0.32\textwidth]{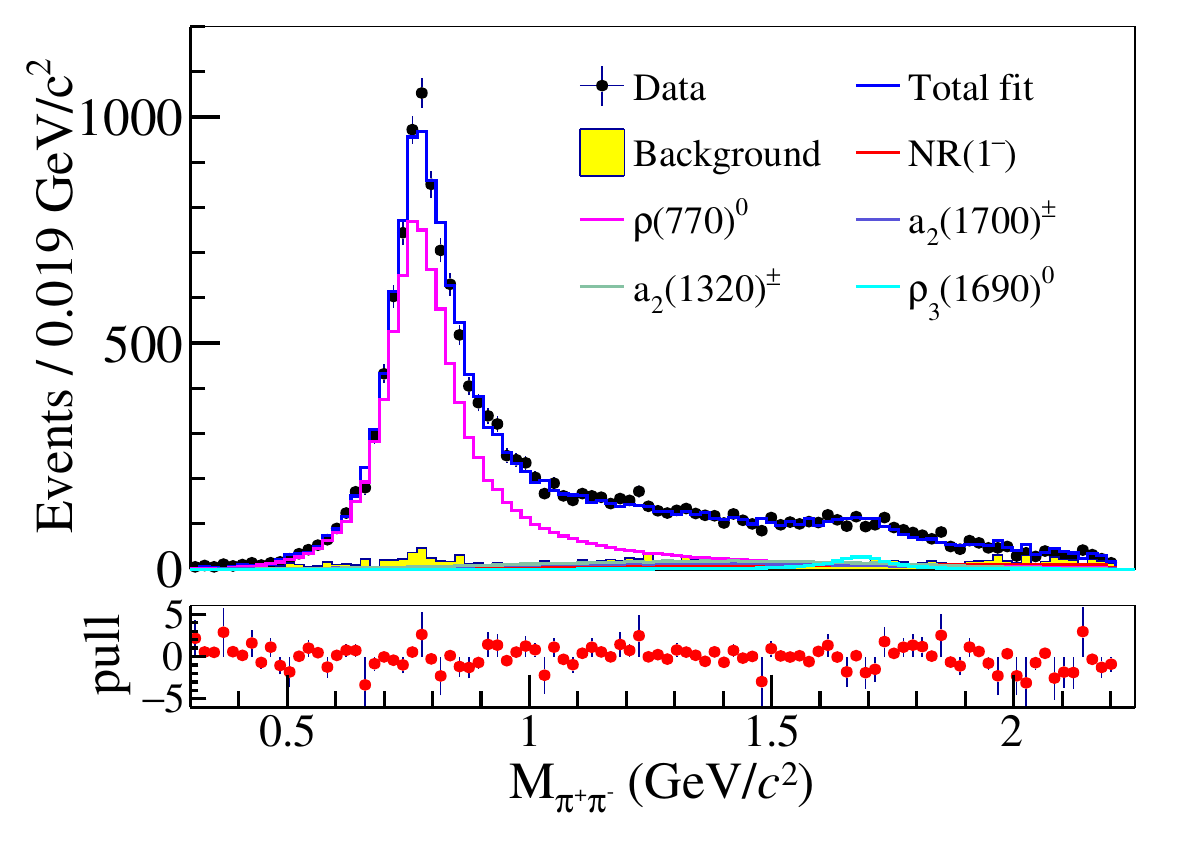}
    \includegraphics[width=0.32\textwidth]{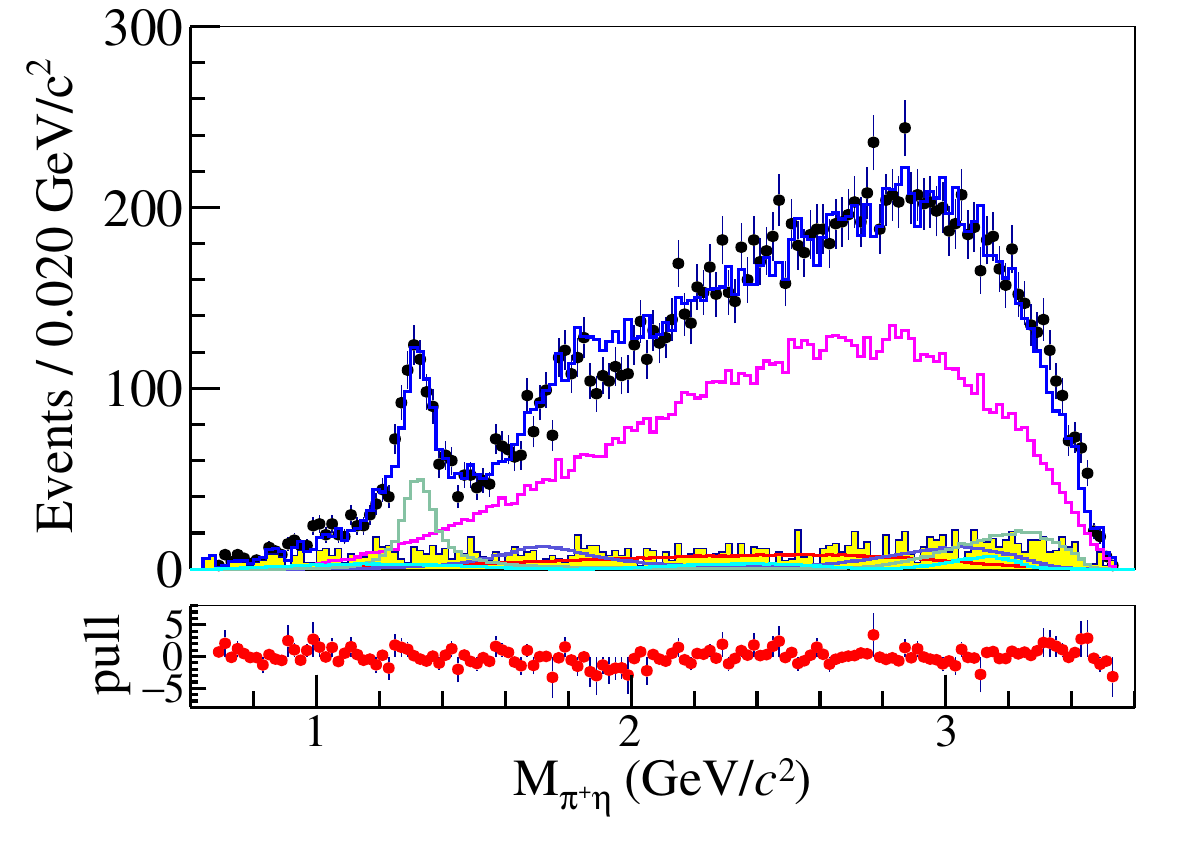}
    \includegraphics[width=0.32\textwidth]{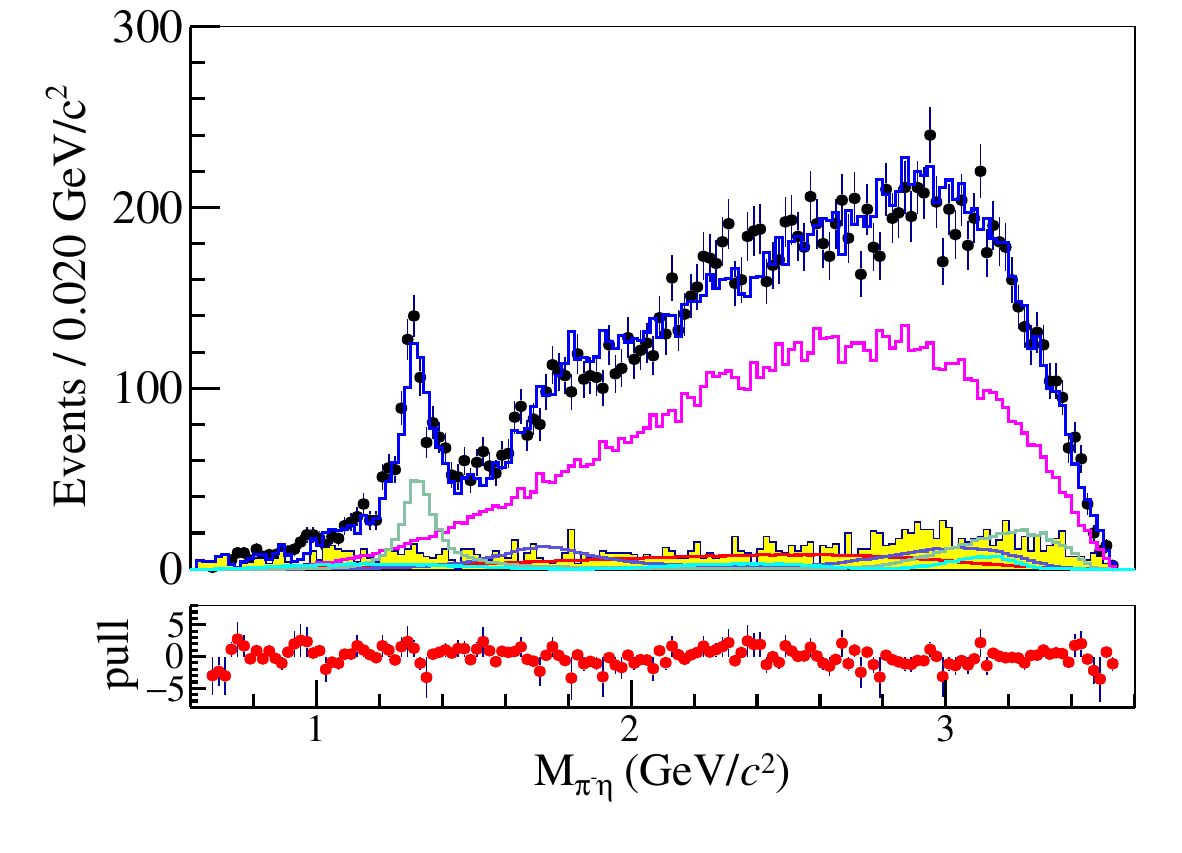}
       \includegraphics[width=0.32\textwidth]{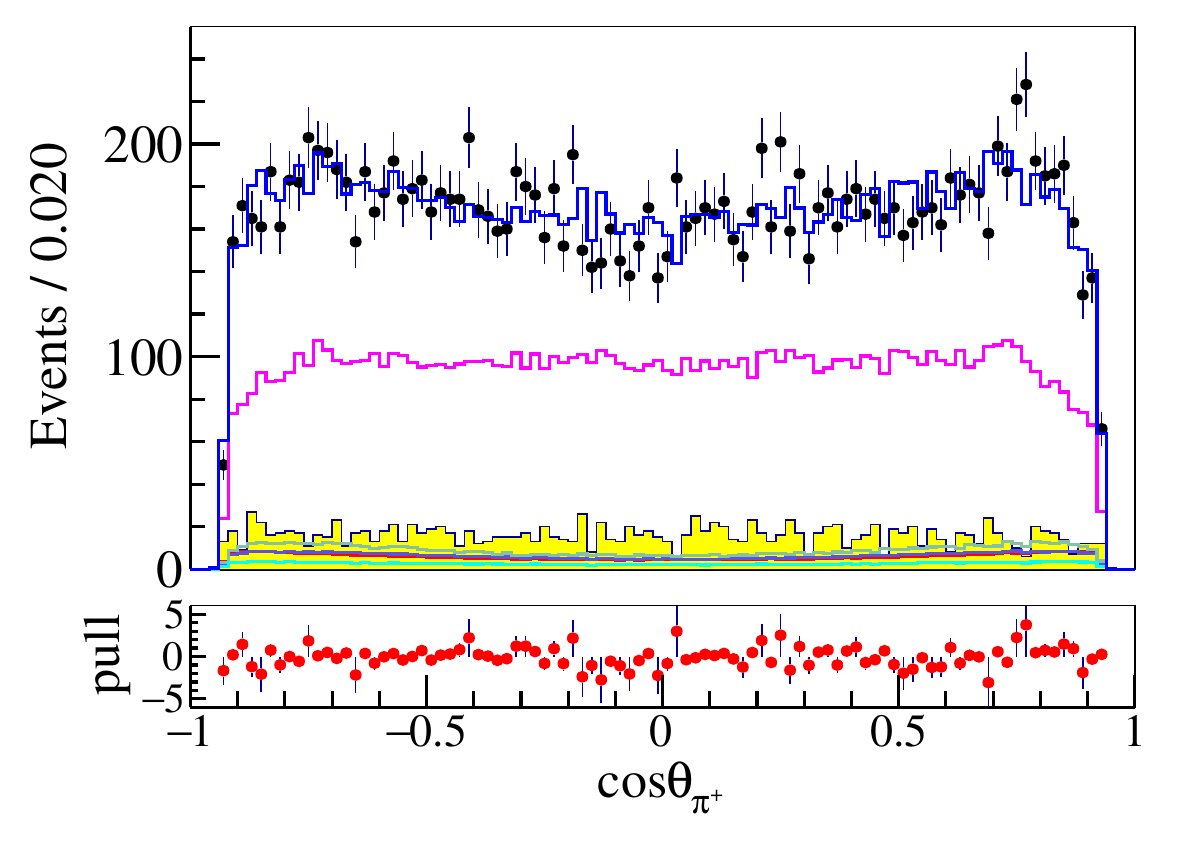}
    \includegraphics[width=0.32\textwidth]{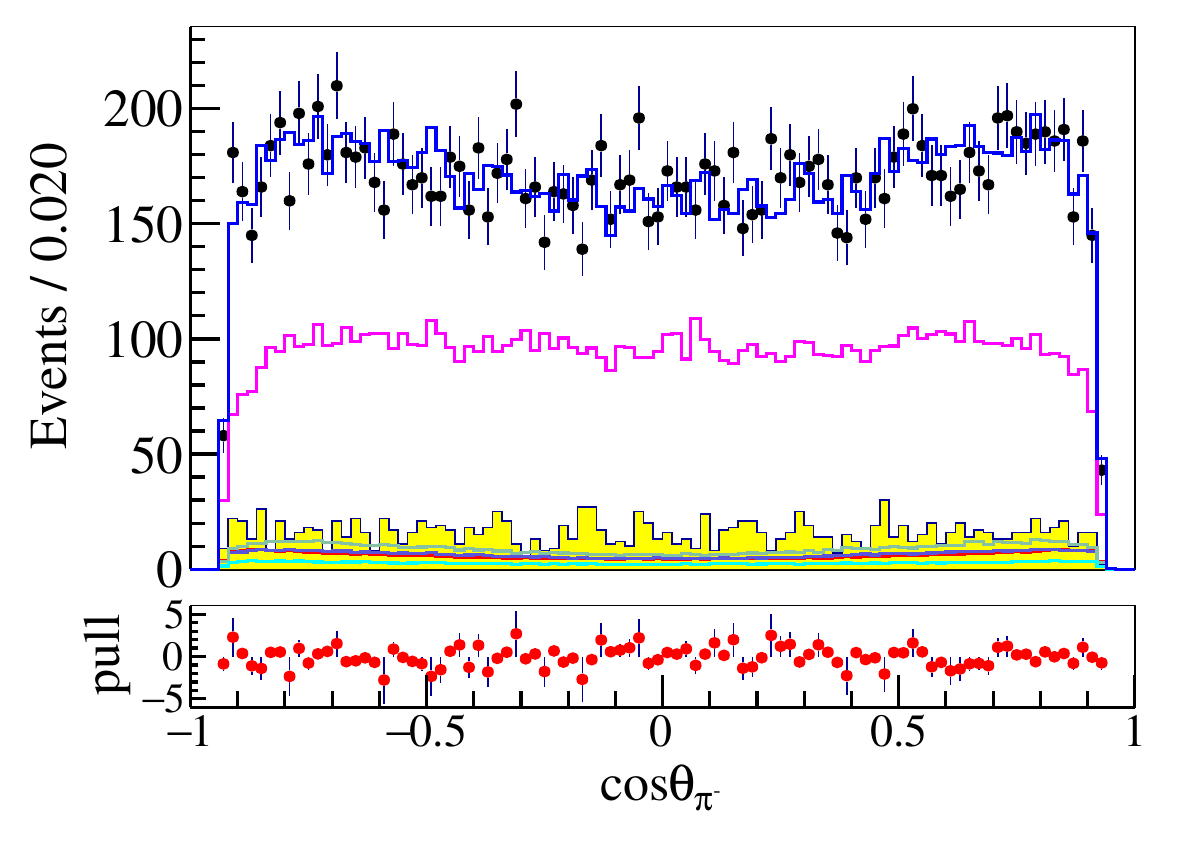}
    \includegraphics[width=0.32\textwidth]{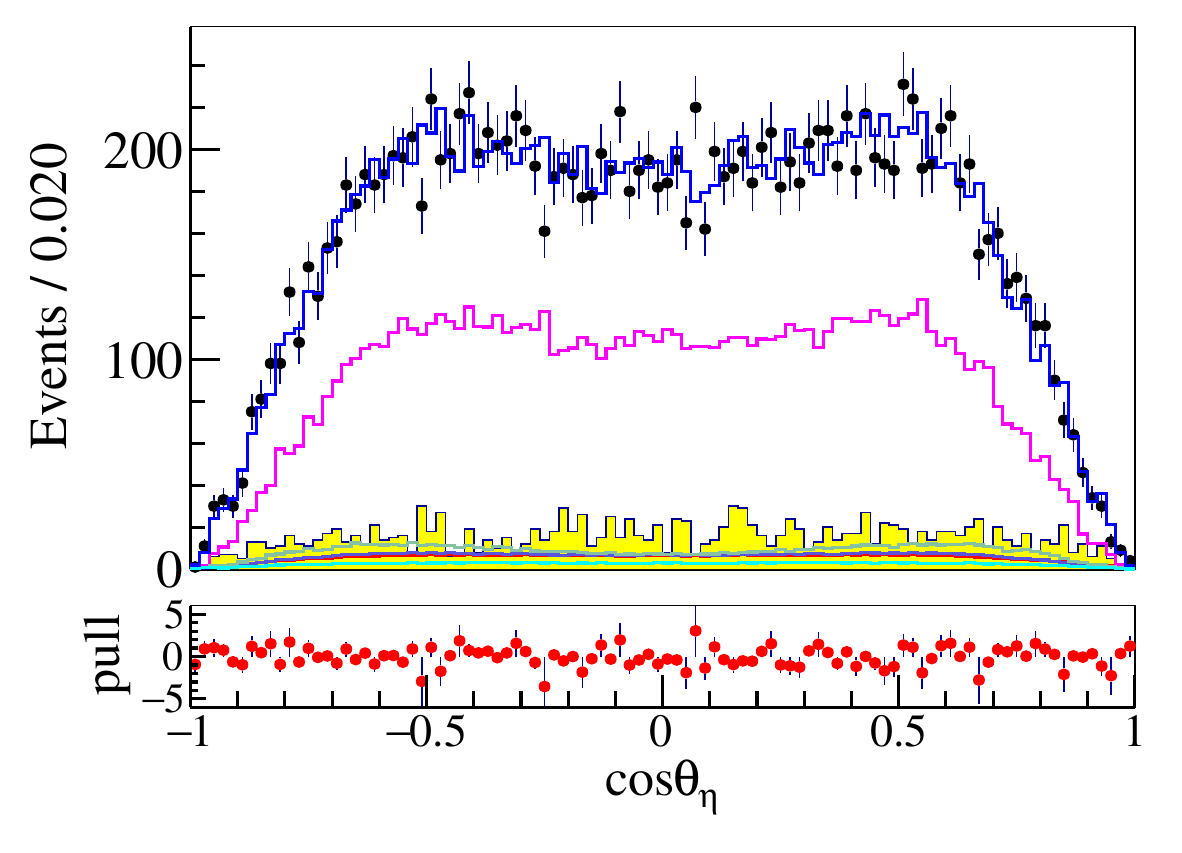}
   \caption{Projections of the PWA fit onto the mass spectra and angular distributions for the data sample collected at $\sqrt{s} = 3.6861$~GeV. The points with error bars denote the data, the solid blue histogram represents the total fit, and the other histograms illustrate the contributions from individual components.}
    \label{fig:pwa_proj}
\end{figure*}

Cross-section fractions and interference contributions for the $\sqrt{s}=3.6861$~GeV dataset are given in Table~\ref{tab:fitfractions}. Statistical uncertainties are determined via the bootstrap method, $\delta R_i=\delta N_i/N_i$. Figure~\ref{fig:pwa_proj} shows projections of the PWA fit onto the $M_{\pi^{+}\pi^{-}}$, $M_{\pi^{+}\eta}$, and $M_{\pi^{-}\eta}$ distributions. The fit projections of the angular distribution are also shown in Fig.~\ref{fig:pwa_proj}, demonstrating good consistency between data and the PWA results.

\begin{table*}[htbp]
    \centering
    \caption{The cross section fractions and interference contributions, in \%, of the components in the nominal fit for the data taken at $\sqrt{s} = 3.6861$
GeV. A negative off-diagonal entry indicates destructive interference. The uncertainties are statistical only.}
    \label{tab:fitfractions}
    \begin{tabular}{lccccccccc}
    \hline\hline
    Mode & NR($1^{-}$) & $\rho(770)^0$ & $a_2(1700)^{\pm}$ & $a_2(1320)^{\pm}$ & $\rho_3(1690)^0$ & & & & \\
    \hline
    NR($1^{-}$) & $3.80\pm0.58$ & $7.79\pm1.09$ & $3.97\pm0.57$ & $1.05\pm0.34$ & $0.01\pm0.00$ & & & & \\
    $\rho(770)^0$   & & $61.60\pm0.83$ & $6.18\pm0.67$ & $5.26\pm0.51$ & $0.07\pm0.01$ & & & & \\
    $a_2(1700)^{\pm}$  & & & $4.12\pm0.42$ & $-2.51\pm0.50$ & $0.40\pm0.18$ & & & & \\
    $a_2(1320)^{\pm}$  & & & & $5.79\pm0.40$ & $0.76\pm0.11$ & & & & \\
$\rho_3(1690)^0$ & & & & & $1.71\pm0.22$ & & & & \\
    \hline\hline
    \end{tabular}
\end{table*}

\section{OBSERVED CROSS  SECTION MEASUREMENTS}

The observed cross section is calculated as:
\begin{equation}
\sigma^{\rm obs} = \frac{N_{\rm sig}}{\mathcal{L} \, \varepsilon \, \mathcal{B}(\eta\to\gamma\gamma)},
\end{equation}
where $N_{\rm sig}$ is the signal yield extracted from the fit to the $M_{\gamma\gamma}$ distribution, and equals to $N_{\rm obs}-N_{\rm bkg}$, $\mathcal{L}$ is the integrated luminosity, and $\varepsilon$ is the detection efficiency determined from the signal MC samples generated according to the PWA results at $\sqrt{s}=3.6861$ GeV.

Because the detection efficiency depends on the input cross section shape through ISR and beam energy spread, we employ an iterative procedure. Initially, we assume a flat cross section  as input to the CONEXC generator. Using the resulting efficiencies, we obtain updated Born cross sections, which are then fed back into CONEXC. The iteration continues until relative efficiency changes become negligible (less than 0.5\%).

The same event selection criteria are applied to the $\psi(3686)$ dataset and the 11 energy-scan points in its vicinity as described in Sec.~III. Table~\ref{tab:crosssections} lists, for each energy point, the observed cross sections for $e^{+}e^{-}\to\pi^{+}\pi^{-}\eta$ with $M_{\pi^+\pi^-}$ in the $\rho(770)^0$ mass region $[0.3, 2.2]~\mathrm{GeV}/c^2$, together with the center-of-mass energies,
integrated luminosities, and detection efficiencies.

\begin{table*}[htbp]
    \centering
    \caption{
    The observed cross sections of $e^{+}e^{-}\to\pi^{+}\pi^{-}\eta$ with   $M_{\pi^+\pi^-}$ in the region $[0.3, 2.2]~\mathrm{GeV}/c^2$ at different $\sqrt{s}$ energy points. Here, $\mathcal{L}$ is the integrated luminosity whose uncertainty is dominated by the systematic effects, $N_{\rm sig}$ is the observed signal yield, $\varepsilon$ is the detection efficiency, 
 and $\sigma^{\rm obs}$ is the observed cross section.  All uncertainties are statistical only.
    }
    \label{tab:crosssections}
    \setlength{\tabcolsep}{8pt} 
\begin{tabular}{c c c c c c}
\hline\hline
$\sqrt{s}$ (MeV) & $\mathcal{L}$ (pb$^{-1}$) & $N_{\rm sig}$ & $\varepsilon$ & $\sigma^{\rm obs}$ (pb) \\ \hline
3581.5 & $84.6$ & $197\pm15$ & 29.6  & $20.0\pm1.5$ \\
3650.0 & $401.0$ & $748\pm29$ & 27.4  & $17.3\pm0.7$ \\
3670.2 & $83.6$ & $148\pm13$ & 27.5 & $16.4\pm1.4$ \\
3680.1 & $83.1$ & $126\pm12$  & 26.3  & $14.6\pm1.4$ \\
3682.0 & $395.6$  & $551\pm25$ & 26.4  & $13.4\pm0.6$ \\
3682.8 & $28.2$  & $39\pm7$  & 26.3  & $13.4\pm2.4$ \\
3684.2 & $27.8$  & $40\pm7$  & 27.8  & $13.1\pm2.3$ \\
3685.3 & $25.3$  & $69\pm9$  & 29.3  & $23.6\pm3.1$ \\
3686.1 & $3877.0$ & $14410\pm131$ & 29.3  & $32.2\pm0.3$ \\
3686.5 & $24.5$  & $102\pm11$  & 30.2  & $35.1\pm3.8$ \\
3691.4 & $68.6$  & $162\pm13$ & 28.3  & $21.2\pm1.7$ \\
3709.8 & $69.3$  & $154\pm13$ & 27.8  & $20.3\pm1.7$ \\ \hline\hline
\end{tabular}
\end{table*}

\section{FIT TO THE cross SECTION LINE SHAPE}

To extract the $\psi(3686)\to\pi^{+}\pi^{-}\eta$ branching fraction while accounting for continuum contributions, we fit the observed cross section line-shape. Because the $\psi(3686)$ resonance is narrow, the observed cross section is significantly affected by the ISR and beam energy spread. 
The process $e^+e^- \to \pi^+\pi^-\eta$ is dominated by the $\rho(770)^0\eta$ intermediate state; this violates isospin and proceeds via electromagnetic interaction.

The Born cross section for $\psi(3686)\to\pi^{+}\pi^{-}\eta$ is given by the coherent sum of the continuum amplitude and the $\psi(3686)$ resonance amplitude~\cite{cross_2,cross_1}:
\begin{equation}
\sigma_{\rm Born}^f(s) = \frac{4\pi\alpha^2}{3s} \left| 1 + \frac{s}{M}\frac{3\sqrt{\Gamma_{ee}^0\Gamma_{\mu\mu}^0}/\alpha}{s-M^2+iM\Gamma}e^{i\phi} \right|^2 \, \mathcal{P}(s) \, \left(\frac{\mathcal{F}}{s^{n/2}}\right)^2,
\end{equation}
where $s$ is the center-of-mass energy squared, $\alpha$ is the fine-structure constant, $\Gamma_{ee}^0$ and $\Gamma_{\mu\mu}^0$ are the bare electronic and muonic widths, $\phi$ is the relative phase between continuum and electromagnetic amplitudes, and $\mathcal{P}(s)$ is the three-body phase-space factor:
\begin{equation}
\mathcal{P}(s) = \frac{1}{32\pi^3 s}\int_{2m_\pi}^{\sqrt{s}-m_\eta} p_1(m) \; p_{12}(s,m) \; dm.
\end{equation}
Here $p_1(m)=\sqrt{(m^2-4m_\pi^2)m^2}/(2m)$ and $p_{12}(s,m)=\sqrt{[s-(m+m_\eta)^2][s-(m-m_\eta)^2]}/(2\sqrt{s})$, with $m$, $m_\pi$, and $m_\eta$ being the $\pi^+\pi^-$ invariant mass, pion, and $\eta$ masses, respectively.

The factor $\mathcal{F}^2/s^n$ is the form factor for hadronization. For a three-body final state, we take $n=5$ following the counting rule in Ref.~\cite{PhysRevD.24.2848}.

The vacuum polarization is included via:
\begin{equation}
\sigma_{\rm dressed}^f (s)= \frac{\sigma_{\rm Born}^f(s)}{|1-\Pi^0(s)|^2},
\end{equation}
where $\Pi^0$ is the vacuum polarization~\cite{Alexander:1988em}. The $\psi(3686)$ resonance contribution is described by a Breit-Wigner function with dressed widths $\Gamma_{ee} = \Gamma_{ee}^0/|1-\Pi^0|^2$ and $\Gamma_{\mu\mu} = \Gamma_{\mu\mu}^0/|1-\Pi^0|^2$, taken from the PDG~\cite{PDG_2024}.

The effect of ISR is incorporated via convolution:
\begin{equation}
\sigma'(s) = \int_{0}^{x_f} dx\, F_{\rm ISR}(x,s)\,\sigma^f_{\rm dressed}(s(1-x)).
\end{equation}
Here, the upper limit of integration is $x_f = 1-s_m/s$, with $s_m$ set at $3.45^2~\mathrm{GeV}^2$.
$F_{\rm ISR}(x,s)$ is the radiator function of Kuraev and Fadin up to $\mathcal{O}(\alpha^2)$~\cite{Kuraev:1985hb,Ping:2013jka}:
\begin{equation}
\begin{split}
  F_{\mathrm{ISR}}(x,s) =&  \, x^{\beta-1}\beta(1+\delta) - \beta\left(1-\frac{x}{2}\right) + \frac{\beta^2}{8} \left[ 4(2-x)\frac{}{} \right. \\
  & \ln\frac{1}{x} \left. - \frac{1+3(1-x)^2}{x}\ln(1-x) - 6 + x \right],
\end{split}
\label{eq:isr_formula}
\end{equation}
where $\delta = \frac{3}{4}\beta + \frac{\alpha}{\pi}\left(\frac{\pi^2}{3}-\frac{1}{2}\right) + \beta^2\left(\frac{9}{32}-\frac{\pi^2}{12}\right)$ and $\beta = \frac{2\alpha}{\pi}\left(2\ln\frac{\sqrt{s}}{m_e}-1\right)$.

Finally, the beam energy spread is accounted for by Gaussian smearing:
\begin{equation}
\sigma_{\rm obs}(s) = \int_{\sqrt{s}-5S_E}^{\sqrt{s}+5S_E} d\sqrt{s'}\,\sigma'(s')\,\frac{1}{\sqrt{2\pi}S_E} \, \exp\left(-\frac{(s-s')^2}{2S_E^2}\right),
\end{equation}
with $S_E = 1.3$ MeV for the $\psi(3686)$ region at BESIII~\cite{corss_3}.

The partial width for $\psi(3686)\to f$ is related to that for $\psi(3686)\to\mu^+\mu^-$ by:
\begin{equation}
\Gamma_f = \mathcal{P}(M^2) \left(\frac{\mathcal{F}}{M^n}\right)^2 \Gamma_{\mu\mu},
\end{equation}
and the branching fraction is:
\begin{equation}
\mathcal{B}(\psi(3686)\to f) = \frac{\Gamma_f}{\Gamma_{\mu\mu}} \, \mathcal{B}(\psi(3686)\to\mu^+\mu^-),
\end{equation}
where  $\mathcal{B}(\psi(3686)\to f)$ is the branching fraction of $\psipipieta$ in the mass region $M_{\pi^+\pi^-}\in [0.3, 2.2]$ ~$\mathrm{GeV}/c^2$.

\begin{table}[htbp]
    \centering
    \caption{Parameters fixed in the cross section line-shape fit.}
    \label{tab:fixedparams}
    \begin{tabular}{c  @{\hspace{3em}}    c}
    \hline\hline
    Parameter & Fixed  \\ \hline
    $M$ (MeV) & 3686.1\\
    $\Gamma$ (keV)& 293.0 \\
    $\mathcal{B}(\psi(3686)\to e^+e^-)$ ($10^{-3}$) & 7.9 \\
    $\mathcal{B}(\psi(3686)\to\mu^+\mu^-)$ ($10^{-3}$) & 8.0 \\
    $n$ &  5\\
    $S_E$ (MeV)& 1.3 \\\hline\hline
    \end{tabular}
\end{table}

The fit is performed with a nuisance parameter treatment of the correlated multiplicative systematic uncertainties~\cite{Cowan_2019,Ball2010}. The $\chi^2$ function is written as
\begin{equation}\label{Equ_chisq}
\chi^2(\vec{\theta}, \eta) = \sum_i  \frac{\left[y_i^{\text{obs}} - (1 + r_i\eta)y_i^{\text{exp}}(\vec{\theta})\right]^2}{\sigma_{i,\text{stat}}^2 + \sigma_{i,\text{uncor}}^2 + \sigma_{i,\sqrt{s}}^2}  + \eta^2,
\end{equation}
where $y_i^{\text{obs}}$ is the measured cross section, $y_i^{\text{exp}}(\vec{\theta})$ is the expected value calculated from the cross section line-shape, and $\vec{\theta}$ represents the fit parameters which describe the line-shape. The parameter $\eta$ is a Gaussian-constrained nuisance parameter, $r_i$ is the relative correlated systematic uncertainty for the $i^{th}$ energy point,  \(\sigma_{i,\text{stat}}\) is the  statistical uncertainty, \(\sigma_{i,\text{uncor}}\) is the point-to-point uncorrelated systematic uncertainty and \(\sigma_{i,\sqrt{s}}\) is the uncertainty associated with the CM energy measurement.

\begin{figure}[htbp]
    \centering    \includegraphics[width=0.47\textwidth]{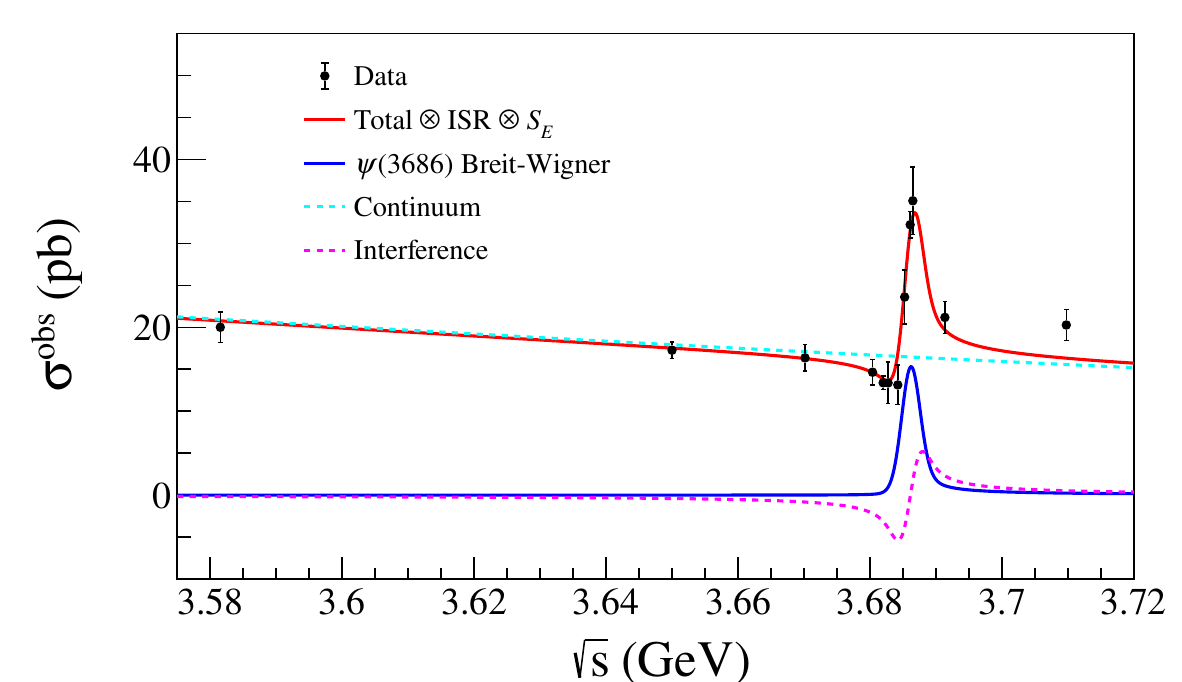}
    \caption{Fit to the observed cross sections of $e^{+}e^{-}\to\pi^{+}\pi^{-}\eta$  with   $M_{\pi^+\pi^-}$ in the region $[0.3, 2.2]~\mathrm{GeV}/c^2$ at different $\sqrt{s}$ energy points.}
    \label{fig:crosssection_fit}
\end{figure}

\begin{table}[htbp]
    \centering
   \setlength{\tabcolsep}{3pt} 
    \caption{Results of the cross section line-shape fit.}
    \label{tab:fitresults}
    \begin{tabular}{ccc}
    \hline\hline
    $\mathcal{B}(\psi(3686)\to\pi^+\pi^-\eta)$ ($10^{-5}$) & $\phi(^\circ)$ & $\chi^2$/ndf \\ \hline
    $2.39\pm0.06\pm0.09$ & $7.76\pm8.08\pm 2.02$ & 9.82/10 \\ \hline\hline
    \end{tabular}
\end{table}

\begin{figure}[htbp]
    \centering
    \includegraphics[width=0.38
    \textwidth]{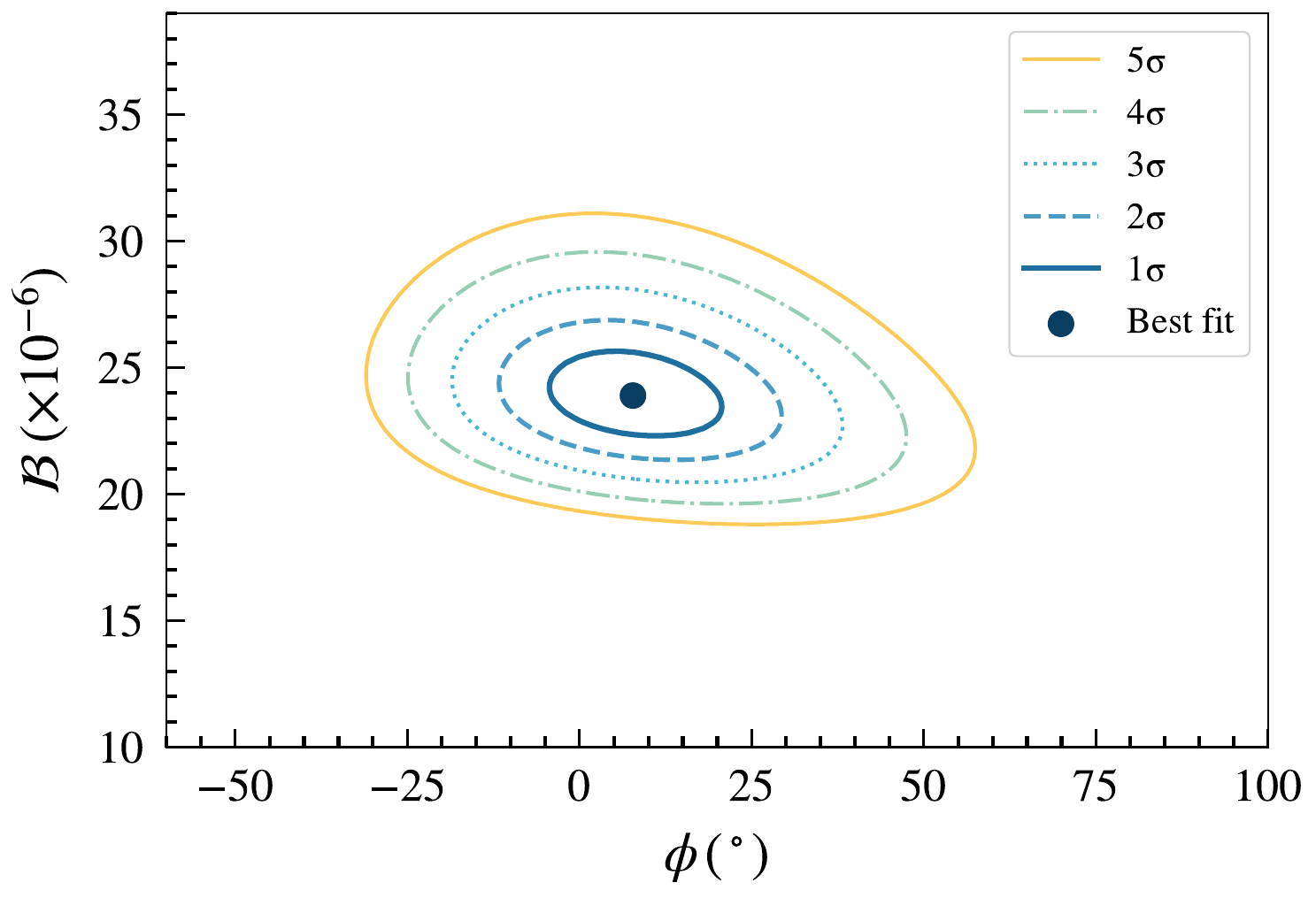}
    \caption{The 2D $\chi^2$ scan vs.~phase $\phi$ and branching fraction $\mathcal{B}(\psi(3686)\to f)$.}
    \label{fig:phasescan}
\end{figure}

The fitting results for the observed cross sections are shown in Fig.~\ref{fig:crosssection_fit}. Fit results for parameters are given in Table~\ref{tab:fitresults}. Figure~\ref{fig:phasescan} shows a $\chi^2$ scan versus the phase $\phi$ and $\mathcal{B}(\psi(3686)\to f)$ ; the $1\sigma$ confidence interval corresponds to $\Delta\chi^2=\chi^2-\chi^2_{\rm min}=2.3$.

\section{SYSTEMATIC UNCERTAINTIES}

\subsection{Cross section measurements}
The systematic uncertainties related to the measurements for cross section are summarized in Table~\ref{tab:crosssys} and described below.
The uncertainty associated with the integrated luminosity is about 1\% at each energy point~\cite{sys_N}. The tracking and PID  efficiencies are all assigned to be 1.0\% per charged track~\cite{sys_track1,sys_track3}, and the photon efficiency is also assigned to be 1.0\% per photon~\cite{sys_photon}. The uncertainty related to the kinematic fit is evaluated via helix correction on charged tracks~\cite{BESIII:2012mpj}. At each energy point, the difference in signal MC reconstruction efficiency between samples with and without the correction is assigned as the corresponding systematic uncertainty. The uncertainty related to $\mathcal{B}(\eta\to\gamma\gamma)$ is taken from the PDG~\cite{PDG_2024}.

Uncertainty from the $M_{\gamma\gamma}$ fit is estimated by varying the fit range by $\pm5$~MeV; the maximum change is assigned as systematic uncertainty. The $M_{\gamma_2}^{\rm rec}$ veto uncertainty is evaluated via a Barlow test~\cite{sys_Barlow}, varying the veto window boundary in 3~MeV steps from 3.440 to 3.470~GeV/$c^2$. The significance $\zeta = |V_{\rm nominal}-V_{\rm test}|/\sqrt{|\sigma^2_{V_{\rm nominal}}-\sigma^2_{V_{\rm test}}|}$ remains below 2 for all variations, so this uncertainty is neglected.

The uncertainty due to the MC modeling is assessed using alternative signal samples that either include the $\omega$ resonance or replace the nonresonant component with a $3^-$ model. The differences in efficiency relative to the nominal sample are taken as the uncertainties. Adding these items in quadrature, we obtain the combined systematic uncertainty for the cross section measurement.

\begin{table*}[htbp]
    \centering
    \caption{Systematic uncertainties in \% on the observed cross sections of $e^{+}e^{-}\to\pi^{+}\pi^{-}\eta$.}
    \label{tab:crosssys}
    \small
    \begin{tabular}{l*{12}{c}}
    \hline\hline
    Source  & 3581.5 & 3650.0 & 3670.2 & 3680.1 & 3682.0 & 3682.8 & 3684.2 & 3685.3 & 3686.1 & 3686.5 & 3691.4 & 3709.8 \\ \hline
    Luminosity & 1.0 & 1.0 & 1.0 & 1.0 & 1.0 & 1.0 & 1.0 & 1.0 & 1.0 & 1.0 & 1.0 & 1.0 \\
    Photon detection & 2.0 & 2.0 & 2.0 & 2.0 & 2.0 & 2.0 & 2.0 & 2.0 & 2.0 & 2.0 & 2.0 & 2.0 \\
    Tracking & 2.0 & 2.0 & 2.0 & 2.0 & 2.0 & 2.0 & 2.0 & 2.0 & 2.0 & 2.0 & 2.0 & 2.0 \\
    PID & 2.0 & 2.0 & 2.0 & 2.0 & 2.0 & 2.0 & 2.0 & 2.0 & 2.0 & 2.0 & 2.0 & 2.0 \\
    Kinematic fit & 0.3 & 0.2 & 0.3 & 0.4 & 0.5 & 0.6 & 0.6 & 0.3 & 0.2 & 0.4 & 0.5 & 0.4 \\
    $\mathcal{B}(\eta\to\gamma\gamma)$ & 0.5 & 0.5 & 0.5 & 0.5 & 0.5 & 0.5 & 0.5 & 0.5 & 0.5 & 0.5 & 0.5 & 0.5 \\
    $M_{\gamma\gamma}$ fit & 3.0 & 1.6 & 1.3 & 1.1 & 0.8 & 0.6 & 1.0 & 0.5 & 2.9 & 0.9 & 0.6 & 1.4 \\
    MC model & 0.2 & 1.1 & 0.3 & 0.1 & 0.9 & 0.6 & 0.1 & 0.6 & 1.1 & 0.5 & 0.1 & 0.5 \\ \hline
    Total & 4.7 & 4.1 & 3.9 & 3.8 & 3.9 & 3.8 & 3.8 & 3.7 & 4.8 & 3.8 & 3.7 & 4.0     
    \\ \hline\hline
    \end{tabular}
\end{table*}

\subsection{Branching fraction measurements}

\begin{table*}[htbp]
    \centering
    \caption{Systematic uncertainties (\%) on branching fraction measurements.}
    \label{tab:brsys}
    \begin{tabular}{lcc}
    \hline\hline
    Source & $\psi(3686)\to\rho(770)^0\eta$ & $\psi(3686)\to a_2(1320)^{\pm}\pi^{\mp}$  \\ \hline
    Cross section measurement & 4.6 & 4.6  \\
    Line-shape fixed parameters & 1.8 & 1.8  \\
    $\mathcal{B}(a_{2}(1320)^{\pm}\to\pi^{\pm}\eta)$ & -- & 8.3  \\
    Intermediate resonance parameters & 0.6 & 6.8  \\
    PWA procedure & 0.5 &2.7   \\
    Background shape & 2.1  &0.3    \\
    Additional resonances & 0.4 & 0.1  \\
    Nonresonant component & 8.7 & 8.6 \\
     \hline
    Total &10.3   &14.9   \\ \hline\hline
    \end{tabular}
\end{table*}

Systematic uncertainties on branching fractions arise from cross section measurements, line-shape fit parameters, and PWA procedures. All the systematic uncertainties on branching fraction measurements are summarized in Table~\ref{tab:brsys}. 

The uncertainties of the line-shape fit due to fixed parameters are evaluated as follows:
the continuum model exponent $n$ is varied by increasing or decreasing it by 1, while $M$, $\Gamma$, 
$\mathcal{B}(\psi(3686) \to e^+e^-)$, and $\mathcal{B}(\psi(3686) \to \mu^+\mu^-)$ 
are varied by $\pm 1$ standard deviation based on PDG values\cite{PDG_2024}. 
The systematic uncertainty on the $\mathcal{B}(a_{2}(1320)^{\pm}\to\pi^{\pm}\eta)$ is  8.3\%~\cite{PDG_2024}.
Systematic uncertainties associated with the PWA are estimated as described below. 

The systematic uncertainties due to  the intermediate resonance parameters, which are taken from the PDG~\cite{PDG_2024} and fixed in the fit, are estimated by performing alternative fits with the parameters varied by $\pm 1\sigma$, and the standard deviation of the resulting yields is taken as uncertainty.
To validate the PWA procedure, an input-output check is done by doing PWAs on 500 sets of toy MC samples to extract the signal yields. The pull is defined as \( P = (V_{\text{out}} - V_{\text{in}})/\sigma_{\text{in}} \), and the corresponding deviations are included in the systematic uncertainties.

To estimate the systematic uncertainty from background determination, alternative fits are performed by varying each background component individually. 
For the peaking background, we change its shape to that obtained from an exclusive MC sample of $\psi(3686) \to  \gamma \pi^+ \pi^- \eta, \eta \to \gamma \gamma$ after applying the same selection criteria. For the non-peaking background, we shift the $\eta$ sideband region to $M_{\gamma\gamma}\in[0.480,0.505]~\mathrm{GeV}/c^2$ in the fit, the difference is taken as the systematic uncertainty.

To evaluate the effects of a possible $\omega$ contribution and its interference with the $\rho(770)^0$, the PWA is re-performed after adding the $\omega$ resonance. The statistical significance of the $\omega$ contribution is found to be $1.2\sigma$. The resulting changes in the fitted fractions are taken as the corresponding systematic uncertainties.
The uncertainty due to the nonresonant  component is evaluated by changing its $J^P$ to $3^-$. Adding these items in quadrature, we obtain the combined systematic uncertainty for the branching fraction measurement.

\section{BRANCHING FRACTIONS AND MIXING ANGLE}

Branching fractions for individual channels are calculated as:
\begin{equation}
\mathcal{B}(\psi(3686)\to X) = \frac{\mathcal{R}_X \, \mathcal{B}(\psi(3686)\to\pi^+\pi^-\eta)}{\mathcal{B}(Y\to\text{final state})},
\end{equation}
where $X$ denotes $\rho(770)^0\eta$ or 
$a_2(1320)^{\pm}\pi^{\mp}$, and $\mathcal{R}_X$ is the 
efficiency-corrected PWA fit fraction determined at 
$\sqrt{s}=3.6861$~GeV. 
We assume that the PWA baseline solution and the continuum--resonance interference pattern are common to both the $\rho(770)^0\eta$ and $a_2(1320)^\pm\pi^\mp$ channels and independent of energy. The data--toy MC agreement at the other energy points indicates that the continuum--resonance interference is consistent with that at $3.6861$~GeV within the current precision.
Here $Y$ denotes $\rho(770)^0$ 
or $a_2(1320)^{\pm}$, and the corresponding secondary branching 
fractions are taken from the PDG~\cite{PDG_2024}.
The results are:
{\small
\begin{align}
\mathcal{B}(\psi(3686)\to\rho(770)^0\eta) &= (1.34 \pm 0.03\pm0.13) \times 10^{-5}, \nonumber\\
\mathcal{B}(\psi(3686)\to a_2(1320)^{\pm}\pi^{\mp}) &= (2.13 \pm0.05\pm0.32 ) \times 10^{-5}.
\end{align}
}
The first uncertainties are statistical and the second systematic.

Using our measured $\mathcal{B}(\psi(3686)\to\rho(770)^0\eta)$ and the two values for $\mathcal{B}(\psi(3686)\to\rho(770)^0\eta^{\prime})$, for either constructive or destructive interference, of $(1.02\pm0.11\pm 0.24)\times 10^{-5}$ or $(0.569\pm0.128\pm 0.236)\times 10^{-5}$ from BESIII~\cite{BESIII_rhoetap}, the $\eta$-$\eta^{\prime}$ mixing angle is obtained with Eq.~\ref{Ratio}. Two solutions are:
\begin{equation}
   \theta_P = (-11.2 \pm 1.6 \pm 3.6)^\circ, \quad
\theta_P = (-19.4 \pm 3.1\pm5.7)^\circ.
\end{equation}
Both solutions agree with the mixing angle from lattice QCD within $1\sigma$~\cite{Christ:2010dd}.

\begin{figure}[htbp]
    \centering
    \includegraphics[width=0.45\textwidth]{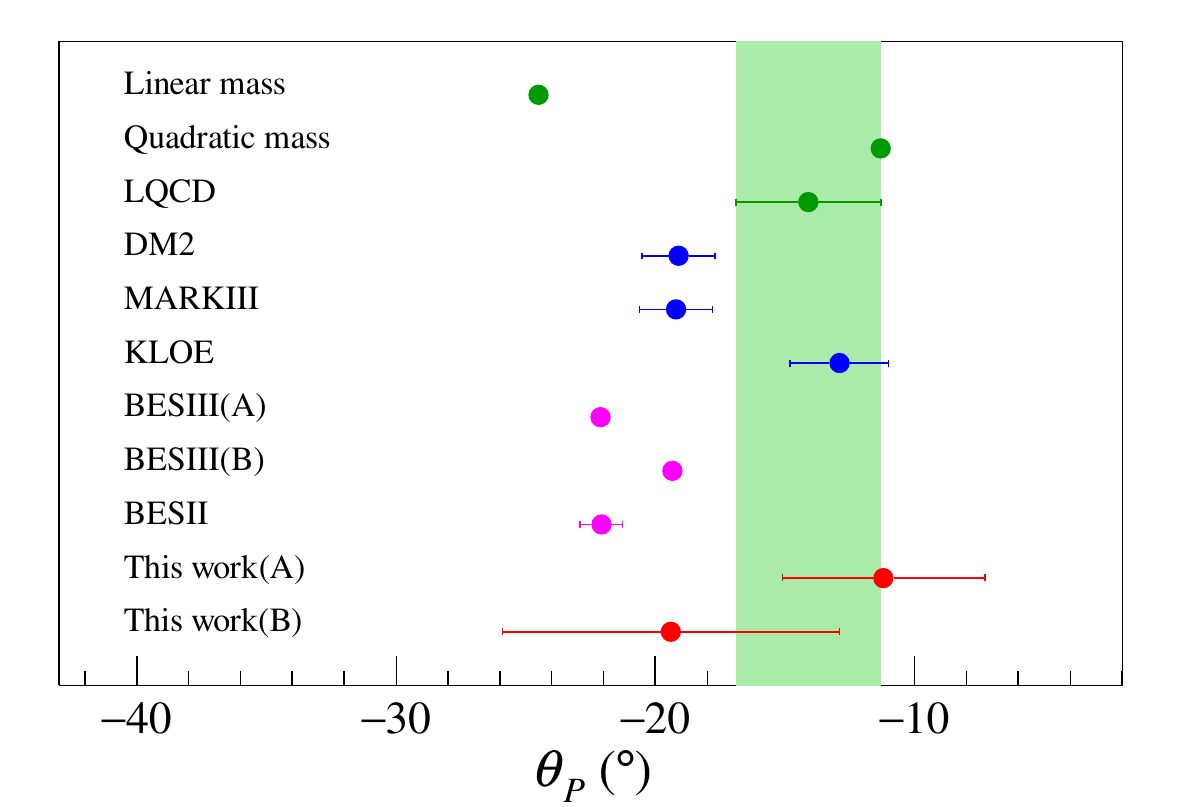}
                \caption{Comparison of determinations of the $\eta$-$\eta^{\prime}$ mixing angle $\theta_P$. The linear and quadratic mass results from Eq.(~\ref{theta_ge}))~\cite{PDG_2024}; LQCD~\cite{Christ:2010dd}, the lattice QCD (LQCD) calculation,  results from DM2~\cite{DM2:1988bfq}, MARKIII~\cite{MarkIII:1988pvy}, KLOE~\cite{KLOE:2002jed}, BESIII(A),(B)~\cite{BESIII:2023fai}, and BESII~\cite{BES:2005otq}, and the results (A), (B)  from this work.}
    \label{fig:mix-angle}
\end{figure}

Figure~\ref{fig:mix-angle} compares the mixing angle $\theta_P$ obtained in this work with previous experimental measurements from DM2, MARKIII, KLOE, BESII and BESIII Collaborations, as well as with lattice QCD theoretical results. Using the PDG inverse-variance weighting method~\cite{PDG_2024}, we combine these experimental values, including the result from this work, with weights determined by the individual uncertainties, $w_i = 1/\delta_i^2$.
The weighted mean and its uncertainty are $\bar{\theta} = \sum w_i \theta_i / \sum w_i$ and $\delta_{\bar{\theta}} = 1/\sqrt{\sum w_i}$. The resulting weighted averages are $(-20.9 \pm 0.2)^\circ$ for solution (A) and $(-21.0 \pm 0.2)^\circ$ for solution (B).

\section{SUMMARY}

Using a sample of $(2712.4 \pm 14.3) \times 10^6$ $\psi(3686)$ events collected with the BESIII detector, we perform a partial wave analysis of $\psi(3686)\to\pi^{+}\pi^{-}\eta$ with the $\pi^{+}\pi^{-}$ invariant mass in the $\rho(770)^0$ mass region $[0.3, 2.2]~\mathrm{GeV}/c^2$. The cross section line-shape around the $\psi(3686)$ resonance is measured with energy-scan data, enabling the estimation of continuum contribution and its interference with the $\psi(3686)$ resonance.

After accounting for interference between the $\psi(3686)$ resonance and the QED continuum, we extract the branching fractions:
\begin{align*}
\mathcal{B}(\psi(3686)\to\rho(770)^0\eta) &= (1.34 \pm 0.03\pm0.13) \times 10^{-5}, \\
\mathcal{B}(\psi(3686)\to a_2(1320)^{\pm}\pi^{\mp}) &= (2.13\pm0.05 \pm 0.32) \times 10^{-5}, 
\end{align*}
where the uncertainties are statistical and systematic, respectively.  The branching fractions $\mathcal{B}(\psi(3686)\to\rho(770)^0\eta)$ is improved compared to previous measurements. 
The decay $\psi(3686)\to a_2(1320)^{\pm}\pi^{\mp}$ is observed for the first time with a significance of $35.7 \sigma$. 

Under the assumption of $\eta$-$\eta^{\prime}$ mixing, the mixing angle is determined to be $\theta_P = (-11.2 \pm 1.6\pm3.6)^\circ$ or $\theta_P = (-19.4 \pm 3.1\pm5.7)^\circ$. 
Both solutions are consistent with lattice QCD expectations within \(1\sigma\)~\cite{Christ:2010dd}. Solution (A) agrees with KLOE measurements~\cite{KLOE:2002jed}, while solution (B) is consistent with all previous results~\cite{DM2:1988bfq,MarkIII:1988pvy,KLOE:2002jed,BES:2005otq,BESIII:2023fai}, providing additional input for studies of pseudoscalar meson mixing.

\textbf{Acknowledgement}

The BESIII Collaboration thanks the staff of BEPCII (https://cstr.cn/31109.02.BEPC) and the IHEP computing center for their strong support. This work is supported in part by National Key R\&D Program of China under Contracts Nos. 2025YFA1613900, 2023YFA1606000, 2023YFA1606704; National Natural Science Foundation of China (NSFC) under Contracts Nos. 
12465015, 12275058, 11635010, 11935015, 11935016, 11935018, 12025502, 12035009, 12035013, 12061131003, 12192260, 12192261, 12192262, 12192263, 12192264, 12192265, 12221005, 12225509, 12235017, 12342502, 12361141819, 12535005; the Chinese Academy of Sciences (CAS) Large-Scale Scientific Facility Program; the Strategic Priority Research Program of Chinese Academy of Sciences under Contract No. XDA0480600; CAS under Contract No. YSBR-101; 100 Talents Program of CAS; The Institute of Nuclear and Particle Physics (INPAC) and Shanghai Key Laboratory for Particle Physics and Cosmology; Agencia Nacional de Investigaci\'on y Desarrollo de Chile (ANID), Chile under Contract No. ANID CCTVal CIA250027; ERC under Contract No. 758462; German Research Foundation DFG under Contract No. FOR5327; Istituto Nazionale di Fisica Nucleare, Italy; Knut and Alice Wallenberg Foundation under Contracts Nos. 2021.0174, 2021.0299, 2023.0315; Ministry of Development of Turkey under Contract No. DPT2006K-120470; National Research Foundation of Korea under Contract No. RS-2026-25486791; National Science and Technology fund of Mongolia; Polish National Science Centre under Contract No. 2024/53/B/ST2/00975; STFC (United Kingdom); Swedish Research Council under Contract No. 2019.04595; U. S. Department of Energy under Contract No. DE-FG02-05ER41374


\bibliography{apssamp}

\end{document}